\documentclass[%
 reprint,
 amsmath,amssymb,
 aps,
]{revtex4-2}

\usepackage{graphicx}
\usepackage{dcolumn}
\usepackage{bm}
\usepackage{mathrsfs}

\begin{document}

\preprint{APS/123-QED}

\title{
Inequality for von Neumann entropy change 
under measurement and dissipation
}

\author{Kohei Kobayashi}
 \altaffiliation[Also at ]{
 Global Research Center for Quantum Information Science, National Institute of Informatics,
\\ 2-1-2, Hitotsubashi, Chiyoda-Ku,Tokyo 101-8430, Japan
 }
 \email{k\_kobayashi@nii.ac.jp}
\affiliation{Global Research Center for Quantum Information Science, National Institute of Informatics,
 2-1-2, Hitotsubashi, Chiyoda-Ku,Tokyo 101-8430, Japan}


\date{\today}

\begin{abstract}

We derive a universal inequality that provides a lower bound on the ensemble-averaged von Neumann entropy change in a quantum system subject to continuous measurement and dissipation. Our result clarifies how entropy production is fundamentally constrained by three distinct contributions: (i) the non-Hermitian structure of the dissipation operator, (ii) the standard variance associated with measurement-induced fluctuations, and (iii) a generalized quantum variance reflecting the noncommutativity between the measurement observable and the quantum state. This third term vanishes when the state and observable commute, and thus represents a purely quantum contribution arising from coherence disturbance and measurement backaction. The derived inequality generalizes classical information-thermodynamic relations, such as the Sagawa–Ueda inequality, to the quantum regime, providing a new perspective on the trade-offs between information acquisition, control, and entropy production in continuously monitored open quantum systems. 

\end{abstract}

\maketitle


\section{Introduction}

The second law of thermodynamics has been deeply reinterpreted from the perspective of information theory.
Recent studies have revealed that acquiring information through measurements and performing subsequent feedback control 
can fundamentally change the limitations imposed by thermodynamic irreversibility \cite{Sagawa1, Sagawa2}.
A prominent example is the Sagawa–Ueda inequality \cite{Sagawa1}, which incorporates the mutual information obtained from measurements into the traditional Clausius inequality and demonstrates that information acts as a thermodynamic resource.
The framework of information thermodynamics has been mainly developed for classical stochastic processes.
However, when extending this framework to quantum systems,
  quantum specific features such as non-commutativity, quantum coherence, and measurement backaction inevitably emerge \cite{Binger}.

In particular, feedback control has recently become powerful tools in experimental quantum technologies.
The measurement-based feedback control enables us to prepare and protect a desired quantum states
 \cite{Wiseman, Stockton, Handel, Jacobs, Geremia, Yanagisawa, Hara, Mirrahimi, Nurdin}.
This protocol extract partial information from a system through continuous measurement and use it to control the dynamics based on measurement outcomes.
Several notable experimental realizations have been reported in recent years \cite{Vijay, Gourgy, Cox}.
Such schemes induce a complex interplay between measurement-induced information gain and environmental dissipation.
Despite their practical importance, a general theoretical framework describing the entropy dynamics under these combined conditions remains underdeveloped.

Entropy in quantum systems is typically characterized by the von Neumann entropy,
which corresponds to the Shannon entropy in classical theory, and whose time evolution reflects non-unitary dissipation \cite{Breuer}.

However, the fundamental principles governing entropy changes under continuous measurement and dissipation remain poorly understood.
In particular, the following two questions are still open:
(i) How does information gained through measurement compete with the irreversibility arising from dissipation?
(ii) How does the non-commutativity between the dissipative operator and its Hermitian conjugate affect entropy production?
Many existing studies focus on specific models or particular continuous measurement schemes \cite{Abe, Toscano, Mohan}, 
but there is a lack of a general inequality formulation and framework that naturally incorporates intrinsic quantum non-commutativity.

In this paper, we present a universal inequality for the von Neumann entropy of a quantum system undergoing continuous measurement and dissipation.
We derive a lower bound on the time derivative of the ensemble-averaged von Neumann entropy, which is expressed as the three distinct contributions:
(i) the expectation value of the commutator between the dissipative operator and its adjoint,
(ii) the measurement-induced variance of the observable, and
(iii) a generalized variance that captures the noncommutativity between the measurement operator and the quantum state.
This result elucidates the thermodynamic role of non-Hermiticity in dissipation and the quantum backaction induced by measurement,
 and gives a fundamental trade-off.
Namely, while measurements extract information and can reduce uncertainty, they also disturb coherence and generate entropy.
Our inequality can be viewed as a quantum version of information-thermodynamic inequality such as the Sagawa–Ueda inequality, connecting classical ideas about information and thermodynamics to quantum systems.

\section{Theoretical framework}

In this section, we introduce the mathematical and physical setting for 
our analysis of entropy dynamics in open quantum systems under continuous measurement and dissipation. 
We begin by formulating the stochastic master equation that describes the evolution of a quantum system subjected to weak continuous measurements, and we clarify the distinction between single-trajectory dynamics and ensemble-averaged behavior.

\subsection{Continuous Quantum Measurement and Feedback}

We consider a quantum system described by a density operator $\rho_t$, 
continuously monitored via an indirect measurement process, such as homodyne detection.
The measurement introduces both information gain and backaction. 
The evolution of the quantum state under such monitoring is given by the stochastic master equation:

\begin{equation}
d\rho_t = -i[u_tH, \rho_t]dt +\mathcal{D}[M]\rho_tdt +\mathcal{D}[L]\rho_tdt+ \mathcal{H}[M]\rho_tdW_t,
\end{equation}
where
\begin{eqnarray}
\mathcal{D}[A]\rho &=&A\rho A^\dagger -\frac{1}{2}A^\dagger A\rho -\frac{1}{2}\rho A^\dagger A, \nonumber \\
\mathcal{H}[A]\rho &=& A \rho + \rho A^\dagger - {\rm Tr}[(A+A^\dagger)\rho] \rho.
\end{eqnarray}
$H$ is the system Hamiltonian and $u_t$ is the time-dependent control input (we set $\hbar=1$ in the following).
$M$ and $L$ are the Lindblad operator representing the interaction between the system and the environment.
$M$ represents the controllable coupling; in the measurement-based feedback control setting, 
$M$ corresponds to the probe for measurement.
On the other hand, $L$ represents the uncontrollable dissipation.
$dW_t=dy_t- {\rm Tr}[(A+A^\dagger)\rho]dt$ is the Wiener increment (standard Brownian motion)
 representing the innovation term based on the measurement outcome $y_t$.
 It has zero mean $\mathbb{E}[dW_t]=0$, and satisfies the Ito rule $dW_tdt=0$ and $dW^2_t=dt$.
This form ensures that the measurement updates the system by using outcomes and induces noise (backaction).

While the stochastic equation describes individual quantum trajectories, 
we are interested in ensemble-averaged quantities such as the average entropy.
The ensemble-averaged density matrix satisfies a deterministic master equation:

\begin{equation}
\label{me1}
\frac{d \mathbb{E}[\rho_t] }{dt} =-i[H, \mathbb{E}[u(\rho_t)\rho_t]] +\mathcal{D}[M]\mathbb{E}[\rho_t] +\mathcal{D}[L]\mathbb{E}[\rho_t].
\end{equation}

Note that Eq. (\ref{me1}) is not linear with respect to $\mathbb{E}[\rho]$.
If $u$ is independent of $\rho$,  Eq. (\ref{me1}) becomes the following master equation:

\begin{equation}
\label{me2}
\frac{d \mathbb{E}[\rho_t] }{dt} =-i[u_t H, \mathbb{E}[\rho_t]] +\mathcal{D}[M]\mathbb{E}[\rho_t] +\mathcal{D}[L]\mathbb{E}[\rho_t].
\end{equation}

\subsection{ Entropy and Its Dynamics}

The central quantity of interest is the von Neumann entropy of the system:
\begin{equation}
S(\rho) = -\mathrm{Tr}[\rho \ln \rho],
\end{equation}
where $0\leq S(\rho_t) \leq \ln d$ with the rank of the quantum system $d$.
$S_t$ takes zero when $\rho_t$ is pure, 
and the maximum is achieved only when the system is maximally mixed $\rho_t=I/d$.
It quantifies the degree of mixedness or uncertainty of the quantum state.
Since the individual are stochastic, we study the ensemble average:
\begin{equation}
\mathbb{E}[S(\rho)] := -\mathbb{E}[\mathrm{Tr}[\rho \ln \rho]].
\end{equation}
This accounts for the entropy of the state over all realizations of the measurement record.
We aim to evaluate and bound the time derivative of $\mathbb{E}[S(\rho_t)] $, 
which captures the rate of information loss or gain in the system under the combined effect of measurement and dissipation.

\section{Main result}

We consider the evolution of the von Neumann entropy of a quantum system 
subject to continuous measurement and dissipation.
The infinitesimal change of the entropy is given by

\begin{eqnarray}
\label{dS}
dS(\rho_t) &=& 
-{\rm Tr}[d \rho_t \ln \rho_t ] - {\rm Tr}[ \rho^{-1}_t (d \rho_t)^2],     \nonumber \\
&=& -{\rm Tr}\left\{ \mathcal{D}[M] \rho_t \ln \rho_t \right\}dt 
- {\rm Tr} \left\{ \mathcal{D}[L] \rho_t \ln \rho_t \right\} dt  \nonumber \\
&\ \ \ \ & - {\rm Tr}\left\{ \rho^{-1}_t (\mathcal{H}[M]  \rho_t )^2 \right\} dt
- {\rm Tr}\left\{  \mathcal{H}[M] \rho_t \ln \rho_t \right\} dW_t, \nonumber \\
\end{eqnarray}
where we have used 
\begin{eqnarray}
&& {\rm Tr}[d \rho_t]=0,  \\ 
&&  {\rm Tr}\left( [H, \rho_t]\ln \rho_t \right)=0.
\end{eqnarray}

Assuming the measurement operator $M$ is Hermitian ($M^\dagger =M$),
 the first term of (\ref{dS}) becomes

\begin{eqnarray}
\label{derive1}
-{\rm Tr}\left\{ \mathcal{D}[M] \rho_t \ln \rho_t \right\}  
&=& {\rm Tr}\left\{ (M^2 \rho_t - M \rho_t M)\ln \rho_t  \right\}. \nonumber \\
\end{eqnarray}

Following the result of \cite{Abe}, this term satisfies the inequality

\begin{eqnarray}
\label{derive2}
- {\rm Tr}\left\{ \mathcal{D}[M] \rho_t \ln \rho_t\right\}
&\geq&  {\rm Tr}\left( [M^\dagger, M] \rho_t \right) \nonumber \\
&=& \langle [M^\dagger, M]\rangle  = 0,
\end{eqnarray}
where $\langle A\rangle ={\rm Tr}[A\rho]$ denotes the quantum expectation value.
This indicates that for Hermitian measurement operators, the measurement-induced Lindblad term 
does not contribute directly to entropy production in this term.
Similarly, the second term in Eq. (\ref{dS}) has the bound:
\begin{eqnarray}
\label{derive3}
-{\rm Tr}\left\{ \mathcal{D}[L]\rho_t \ln \rho_t\right\} 
 \geq  \langle [L^\dagger, L]\rangle.
\end{eqnarray}

This term reflects entropy production due to irreversible dissipative processes, 
which arises from the non-Hermiticity of the operator $L$ and reflects the asymmetry in energy flow (e.g., between absorption and emission).

The third term of (\ref{dS}) which stems from the stochastic measurement backaction, can be expanded as follows:

\begin{eqnarray}
\left(\mathcal{H}[M]\rho_t \right)^2
&=& \left(M\rho_t +\rho_t M-2{\rm Tr}[M \rho_t]\rho_t \right)^2  \nonumber   \\
&=&M \rho_tM \rho_t +  M\rho_t^2 M + \rho_t M^2 \rho_t + \rho_t M \rho_t M  \nonumber   \\
&\ \ \ & -2{\rm Tr}[M \rho_t](\rho_t M\rho_t +\rho_t^2 M )  \nonumber   \\
&\ \ \ &-2{\rm Tr}[M \rho_t](\rho_tM \rho_t + M\rho_t^2 )   
+4{\rm Tr}[M \rho_t]^2.  \nonumber   \\
\end{eqnarray}

After simplification, the contribution of this term to the entropy change becomes:

\begin{eqnarray}
\label{derive4}
&-& {\rm Tr}\left\{ \rho^{-1}_t (\mathcal{H}[M]  \rho_t )^2\right\}  \nonumber   \\
&=& -3 {\rm Tr}[M^2 \rho_t] + 4{\rm Tr}[M\rho_t ]^2-{\rm Tr}[\rho^{-1}_t M\rho_t^2 M ]  \nonumber   \\
&=& -3{\rm Var}[M] - {\rm Var}^{\rm gen}[M],
\end{eqnarray}

where we define the standard variance of $M$ as
\begin{eqnarray}
{\rm Var}[M] := {\rm Tr}[M^2 \rho_t] - {\rm Tr}[M \rho_t]^2,
\end{eqnarray}
and the generalized variance as
\begin{eqnarray}
{\rm Var}^{\rm gen}[M] := {\rm Tr}[\rho_t^{-1} M \rho_t^2 M] - {\rm Tr}[M \rho_t]^2.
\end{eqnarray}

This generalized variance quantifies how the noncommutativity between the state $\rho_t$ and the observable $M$
 enhances entropy production beyond classical fluctuations.
When $[M, \rho_t] = 0$, the expression reduces to ${\rm Var}[M]$.
However, when $M$ and $\rho_t$ do not commute, ${\rm Var}^{\rm gen}[M]$
represents the coherence disturbance caused by measurement in a basis misaligned with the eigenbasis of the state.

By combining the above Eqs, we obtain the following inequality for the infinitesimal change of entropy:

\begin{eqnarray}
\label{dSdt}
dS(\rho_t) &\geq&
\langle [L^\dagger, L] \rangle -3 {\rm Var}[M]-{\rm Var}^{\rm gen}[M] \nonumber   \\
&\ \ \ & -{\rm Tr}\left\{ \mathcal{H}[M] \rho_t \ln \rho_t \right\} dW_t.
\end{eqnarray}

Taking the ensemble average over measurement trajectories (i.e., the stochastic average over $W_t$), 
the last term vanishes because $\mathbb{E}[dW_t]=0$. 
Therefore, we arrive at the following bound on the average rate of entropy change:

\begin{eqnarray}
\label{result}
\frac{d \mathbb{E}[ S_t] }{dt} \geq
\mathbb{E}[ \langle [L^\dagger, L] \rangle ]-3 \mathbb{E}[ {\rm Var}[M] ]-\mathbb{E}[ {\rm Var}^{\rm gen}[M] ].  \nonumber   \\
\end{eqnarray}

This inequality provides a lower bound on the rate of von Neumann entropy change in an open quantum system 
undergoing both continuous measurement and dissipation.  
It highlights how entropy production is fundamentally constrained by three quantum features:  
(i) the non-Hermiticity of the dissipative operator,  
(ii) the statistical uncertainty due to measurement, and  
(iii) the noncommutativity between the quantum state and the measurement observable.

The first term on the righthand side of Eq. (\ref{result}) reflects the degree of non-Hermiticity of the dissipative operator $L$, 
corresponding to the irreversibility of the quantum dynamics due to information leakage into environments.  
When $L$ is non-Hermitian, the commutator $[L^\dagger, L]$ becomes nonzero and quantifies asymmetry in the dissipative processes, 
such as the imbalance between excitation and relaxation.  

The second and third terms corresponds to the contribution of the measurement.  
The second term, involving the variance of $M$, accounts for classical-like statistical fluctuations due to measurement.  
The third term, ${\rm Var}^{\rm gen}[M]$, represents the genuinely quantum contribution; 
how the noncommutativity between $\rho$ and $M$ induces entropy production through quantum backaction and coherence disturbance.

This result establishes a trade-off between information gain and quantum disturbance, 
and can be viewed as a quantum generalization of information-thermodynamic inequalities such as the Sagawa–Ueda inequality.  
It reveals how entropy generation in continuously monitored quantum systems is governed not only by classical uncertainty but also by intrinsic quantum incompatibility.

Here, we particularly focus on a special case when the dissipative operator is Hermitian, i.e., $L=L^\dagger$.
In this situation, the commutator term $\langle [L^\dagger, L]\rangle$ vanishes identically, and our inequality reduces to

\begin{eqnarray}
\frac{d\mathbb{E}[S_t] }{dt} \geq -3\mathbb{E}[ {\rm Var}[M]]
-\mathbb{E}[ {\rm Var}^{\rm gen}[M]].
\end{eqnarray}

The inequality reveals that the entropy increase is solely governed by the effect of measurement.
The righthand side describes how the noncommutativity of measurement operators with the quantum state induces entropy production.
This provides a fundamental lower bound on the thermodynamic cost of quantum measurement;
 acquiring information through measurement inevitably disturbs the system and increases entropy, even without external dissipative influence.

\subsection{Relation to the Sagawa-Ueda Inequality}

Our result can be viewed as a quantum and entropic extension of the Sagawa-Ueda inequality \cite{Sagawa1}:

\begin{eqnarray}
W_{\rm ext} \geq \Delta F-k_B T I(\rho),
\end{eqnarray}
where $W_{\rm ext}$ is the work extracted via feedback control, $\Delta F$ is the free energy difference,
$k_{B}$  is the Boltzman constant, $T$ is the temperature of the bath,
 and $ I(\rho)$ denotes the mutual information acquired through measurement. 
This inequality relates information gain through measurement to thermodynamic work in systems under feedback control. 
The original Sagawa-Ueda framework focuses on the modification of the second law via mutual information and free energy differences.
Our inequality addresses the dynamical behavior of the von Neumann entropy under continuous measurement and dissipation.
In particular, we demonstrate that the time derivative of the average von Neumann entropy is bounded from 
below by the difference between the measurement-induced variance and the average of a commutator characterizing dissipation.
This formulation makes the thermodynamic role of quantum measurement explicit and clarifies how strong measurement can suppress entropy production and stabilize the system.
Thus, our result generalizes the insight of Sagawa and Ueda by focusing not on extractable work but on the entropic structure of open quantum dynamics under measurement.

\section{Conclusion}

In this paper, we have derived a universal lower bound on the ensemble-averaged rate of von Neumann entropy change in a 
quantum system governed by a stochastic master equation, which incorporates both continuous measurement and dissipative dynamics.
The resulting inequality reveals that entropy production is fundamentally constrained by three key contributions:
 (i) the non-Hermitian structure of the dissipation operator, 
 (ii) the variance of the measurement observable, 
 and (iii) a generalized variance term that arises from the noncommutativity between the quantum state and the measurement operator.

The inequality reveals that the entropy production is mainly constrained from below by two distinct contributions: the non-Hermitian structure of the dissipation operator, and the measurement-induced fluctuations within the system.
The first term, $\langle [L^\dagger, L]\rangle$, reflects the intrinsic irreversibility associated with dissipation channels.
The second term, the variance of the measurement operator, quantifies the information gain and backaction from the measurement process.
The third term, the generalized variance ${\rm Var}^{\rm gen}[M]$, arises exclusively from quantum noncommutativity 
and reflects how measurements disturb the coherence of the quantum state.
Notably, this term becomes ${\rm Var}[M]$ when the measurement operator commutes with the quantum state.
These contributions provide a new thermodynamic perspective on the cost of feedback and control in open quantum systems, 
generalizing the classical Sagawa–Ueda inequality to the regime of quantum trajectories and continuous observation.
Our result offers a theoretical foundation for analyzing entropy dynamics in non-equilibrium quantum systems, 
and may be help for the design of entropy-efficient quantum control protocols, with potential applications in quantum information processing and quantum thermodynamic devices.

\begin{acknowledgments}
This work was supported by MEXT Quantum Leap Flagship Program Grant JPMXS0120351339.
\end{acknowledgments}


\end{document}